# Identifying the Importance of Software Reuse in COCOMO81, COCOMOII.

CH.V.M.K.Hari[#1]     Prof. Prasad Reddy P.V.G.D[$2]     J.N.V.R Swarup Kumar[*3]     G.SriRamGanesh[*3]

[#1]*Associate Professor, Dept of IT, Gitam University, Visakhapatnam, India,* kurmahari@gmail.com.
[$2] *Dept.of CS & SE, Andhra University, Visakhapatnam, India.* prof.prasadreddy@gmail.com
[*3] *Dept. of IT, Gitam University, Visakhapatnam, India,*
swarupjnvr@yahoo.co.in, sriramganesh_g@yahoo.co.in.

*Abstract-* **Software project management is an interpolation of project planning, project monitoring and project termination. The substratal goals of planning are to scout for the future, to diagnose the attributes that are essentially done for the consummation of the project successfully, animate the scheduling and allocate resources for the attributes. Software cost estimation is a vital role in preeminent software project decisions such as resource allocation and bidding. This paper articulates the conventional overview of software cost estimation modus operandi available. The cost, effort estimates of software projects done by the various companies are congregated, the results are segregated with the present cost models and the MRE (Mean Relative Error) is enumerated. We have administered the historical data to COCOMO 81, COCOMOII model and identified that the stellar predicament is that no cost model gives the exact estimate of a software project. This is due to the fact that a lot of productivity factors are not contemplated in estimation process. The vital dilemma we identified is that "software reuse" is being eclipsed although most of the contemporary software projects are based on object oriented development where no component is made from scratch (Inheritance). By using the principal of software reuse the ROI (Return of Investment) is also bolstered for the companies. So further research exposure is in "software Reuse" and Reuse software cost estimation model.**

I. INTRODUCTION

The concept of software cost estimation has been growing rapidly due to practically and demand for it. Today the people expecting high quality of software with a low cost that is goal of software engineering. So many popular cost estimation models like COCOMO81, COCOMOII, SLIM, FP and Delphi. These models created by taking historical data applied to regression analysis. A recent review of surveys on software cost estimation found that of software projects have cost overruns. Today most of the software companies follow COCOMOII for estimating the cost of products; we found some variations in this model [11]. These are several reasons like "unrealistic over-optimum", "complexity", "and overlooked tasks" [9]. The reason we identified are the people are developing the projects by using Object Oriented Technologies with the principle of "software Reuse". This paper we are present some popular software cost estimation models and applied sample data to models and calculated the MRE. In section2 deals with the overview of the cost estimation models. In section5 deals with the calculation by using COCOMO81, COCOMOII and comparison graphs for COCOMO models.

The contribution of this paper predicts the importance the "Software Reuse". Cost Estimation process is an uncertain activity because of inaccurate information and future needs are not known in advance.

II. BACKGROUND

A review of the literature tells the most interesting difference between estimated effort and original effort, estimation models that use KDLOC (Thousands of Delivered Lines of Code) as the primary input. This input is not sufficient for accurately estimating the cost of products. Several other parameters have to be considered. We examine the COCOMO81, COCOMOII models. After examining these models we found some variations in these models. We identified the large scale reuse offered by product line engineering promises a best productivity and time-to-market.

*A. Single variable method*

Software cost estimation is the method for analyzing and predicting the amount of effort required to build a software system. A traditional approach to estimate effort of software creation and development is to make the effort as the function of a single variable. The variable which we use in this model is project size [4].

$$Effort = a * size^b$$

Where effort is in person-months, a & b are constants determined by regression analysis applied on historical data.

*B. COCOMO81 Model*

Boehm described COCOMO as a collection of three variants: basic model, intermediate model, detailed model [12].

*1) Basic model*

The basic COCOMO model computes effort as function of program size, and it is same as single variable method.

$$Effort = a * size^b$$

Where a and b are the set of values depending on the complexity of software. For the organic type of projects a=2.4, b=1.05, semi-detached type of projects a=3.0, b=1.12, Embedded type of projects a=3.6, b=1.2.

*2) Intermediate model*

An intermediate COCOMO model effort is calculated using a function of program size and set of cost drivers or effort multipliers.

$$\text{Effort} = (a*size^b)*EAF$$

where a and b are the set of values depending on the complexity of software and EAF (Effort Adjustment Factor) which is calculated using 15 cost drivers [12]. Each cost driver is rated from ordinal scale ranging from low to high. For the organic type of projects a=3.2, b=1.05, semi-detached type of projects a=3.0, b=1.12, Embedded type of projects a=2.8, b=1.2.

*3) Detailed model*

In detailed COCOMO the effort is calculated as function of program size and a set of cost drivers given according to each phase of software life cycle. The phases used in detailed COCOMO are requirements planning and product design, detailed design, code and unit test, and integration testing.

$$\text{Effort} = (a*size^b)*EAF*sum(Wi).$$

The weights of life cycle model are described in [12]. The life cycle activities like requirement planning, system design, detailed design, code and unit testing, integration and testing. In all above three models the factors a and b are depend on the development mode.

*C. COCOMO II Model*

Boehm and his colleagues have refined and updated COCOMO called as COCOMO II.
This consists of application composition model, early design model, post architecture model.

*1) The Application Composition Model*

It uses object points for sizing rather than the size of the code. The initial size measure is determined by counting the number of screens, reports and the third generation components that will be used in application.

$$\text{Effort} = NOP/PROD$$

Where NOP (New Object Points) = (object points)*(100-%reuse)/100, PROD (Productivity Rate)=NOP/PersonMonths

*2) The Early Design Model*

It uses to evaluate alternative software system architectures where unadjusted function point is used for sizing.

$$\text{Effort} = a*KLOC*EAF$$

Where a is set to 2.45, EAF is calculated as in original COCOMO model using seven cost drivers (RCPX, RUSE, PDIF, PERS, PREX, FCIL, SCED) [12]. RUSE: Reuse is consider as one factor, but it is a major factor for effort estimation.

*3) The Post Architecture Model*

It is used during the actual development and maintenance of a product. The post architecture model includes a set of 17 cost drivers [12] and a set of 5 factors determining the projects scaling component.

$$\text{Effort}=(a*size^b)*EAF$$

Where a=2.55 and b is calculated as b=1.01+0.01*SUM(wi), wi= sum of weighted factors.

*D. SLIM Model*

Larry Putnam of Quantitative Software Management developed The Software Lifecycle Model (SLIM) in 1970's [1,2,11]. SLIM is based on the concept of Norden-Rayleigh curve which represents manpower as a function of time. The software equation for SLIM is defined as follows:

$$S = E*(\text{Effort})^{1/3}*td^{4/3}$$

Where td is the software delivery time, E is the environment factor that reflects the development capability, which can be derived from historical data using the software equation. The size S is in LOC and the Effort is in person-year. Another important relation is

$$\text{Effort} = D_0*td^3$$

Where $D_0$ is a parameter called manpower build-up which ranges from 8 (entirely new software with many interfaces) to 27 (rebuilt software). Combining the above equation with the software equation, we obtain the power function form:

$$\text{Effort} = (D_0^{4/7}*E^{-9/7})*S^{9/7} \text{ and}$$
$$td = (D_0^{-1/7}*E^{-3/7})*S^{3/7}$$

SLIM is widely used in practice for large projects (more than 70 KDLOC) and SLIM is a software tool based on this model for cost estimation and manpower scheduling.

*E. Function Point Analysis (FP)*

It is one of the major techniques used for software cost estimation. It was introduced by Albertch [11].

The general approach that FPA follows is

- Count the number of inputs, outputs, inquiries, master files, and interfaces required, then calculate the Unadjusted Function Points (UFP)

- Calculate the adjusted function point (AFP) by multiplying these counts by an adjustment factor; the UFP and the product complexity adjustment.

- Calculate the Source Lines of Code (SLOC) with the help of the AFP and the Language Factor (LF).

The FPA measures functionality that the user requires like the number of inputs, outputs, inquiries, master files, and interfaces required. The specific user functionality is a measurement of the performance delivered by the application as per the request of the user. For each function identified above the function is further classified as *simple, average* or *complex* and a weight are given to each. The sum of the weights quantifies the size of information processing and is referred to as the Unadjusted Function points. The function types and the weighting factors for the varying complexities [11].

To calculate the Complexity adjustment value, several factors have to be considered, such as Backup and recovery, code design for reuse, etc. All the factors and their estimated values in this project are already available. The adjusted function point denoted by FP is given by the formula:

FP = total UFP*(0.65 + (0.01 *Total complexity adjustment value)) or
FP =total UFP *(Complexity adjustment factor)

Total complexity adjustment value is counted based on responses to questions called complexity weighting factors [11,12]. Each complexity weighting factor is assigned a value (complexity adjustment value) that ranges between 0 (not important) to 5 (absolutely essential).

*F. Delphi model*

This model also known as an expert judgment model, this model has been followed by most of the software companies that we have observed in literature survey. A meeting has been conducted for the experts and predicting the requirements about the project and collect the estimations from all experts and distribute to all of them for discussion and finally and the cost is determined by the following formula

Estimation=(leastestimation+4*avgestimation+highestimation)/6

### III. RESEARCH QUESTIONS

This copious study of software cost estimation ruminates contemporary cost estimation models and tries to contemplate on the differences that prevail between original effort and calculated effort. It also considers manifold cases and tabularizes them in an elucidatory manner. The main models that we scrutinize are the COCOMO, Function Point model and SLIM.

Q1: Why does a discrepancy arise between the original effort and calculated effort? What are the factors that are being precluded by the user while gauging the cost?

Q2: Which factor portrays a vital role in software development and would reduce the difference between actual effort and calculated effort?

### IV. SURVEY METHOD

This research has progressed by excogitating on the famous cost estimation models in hope of unveiling the different ways of guesstimating the cost for a software project. The formulae from the various books, web and journals have been congregated and also historical data from past projects has been collected. The parameters which have been deliberated are based on regression analysis of the different models. We have amassed data from 30 projects [11,23,24] done by renowned companies and this data has been exercised on all the models and MRE has been calculated. This exhibits a lot of difference between actual effort and calculated effort in various models. Based on our astute observation there is no commodious cost estimation model that dispenses with manifold projects. We have visited personnel working with acclaimed organizations and enquired them in order to find evidence and most companies follow expert judgment for determining the cost of the product. Some have admitted that they use a lot of software tools for developing the product and construct programs from existing libraries.

### V. PRELIMINARY RESULTS AND FUTURE RESEARCH (PERFORMANCE OF ESTIMATION MODELS)

TABLE I
COCOMO81 Basic Model

| References | project no. | size(kloc) | Original Effort | Organic(1.05) Basic Effort | Error % | Semidetached(1.12) Basic Effort | Error % | Embedded(1.20) Basic Effort | Error % |
|---|---|---|---|---|---|---|---|---|---|
| * | 1 | 50 | 47 | 145 | 208 | 239 | 408 | 393 | 736 |
| * | 2 | 40 | 66 | 115 | 74 | 186 | 181 | 301 | 356 |
| * | 3 | 22 | 66 | 61 | -7 | 95 | 43 | 146 | 121 |
| * | 4 | 13 | 159 | 35 | -77 | 53 | -66 | 78 | -50 |
| * | 5 | 12 | 218 | 32 | -85 | 48 | -77 | 71 | -67 |
| * | 6 | 34 | 723 | 97 | -86 | 155 | -78 | 247 | -65 |
| * | 7 | 6.2 | 775 | 16 | -97 | 23 | -97 | 32 | -95 |
| * | 8 | 2.5 | 312 | 6 | -98 | 8 | -97 | 10 | -96 |
| * | 9 | 5.3 | 883 | 13 | -98 | 19 | -97 | 26 | -97 |
| * | 10 | 19.5 | 433 | 54 | -87 | 83 | -80 | 127 | -70 |
| * | 11 | 28 | 337 | 79 | -76 | 125 | -62 | 196 | -41 |
| * | 12 | 30 | 345 | 85 | -75 | 135 | -60 | 213 | -38 |
| * | 13 | 32 | 302 | 91 | -69 | 145 | -51 | 230 | -23 |
| * | 14 | 57 | 452 | 167 | -63 | 277 | -38 | 460 | 1 |
| ** | 15 | 30.8 | 143.7 | 87 | -39 | 139 | -3 | 220 | 53 |
| ** | 16 | 34.8 | 161.3 | 99 | -38 | 159 | -1 | 254 | 57 |
| ** | 17 | 38.8 | 178.6 | 111 | -37 | 180 | 0 | 290 | 62 |
| *** | 18 | 39 | 542 | 112 | -79 | 181 | -66 | 292 | -46 |
| **** | 19 | 128.6 | 557 | 393 | -29 | 690 | 23 | 1222 | 119 |
| **** | 20 | 15.4 | 400 | 42 | -89 | 64 | -84 | 95 | -76 |
| **** | 21 | 11.3 | 240 | 30 | -87 | 45 | -81 | 66 | -72 |
| **** | 22 | 12.3 | 95 | 33 | -65 | 49 | -48 | 73 | -23 |
| **** | 23 | 13.3 | 87 | 36 | -58 | 54 | -37 | 80 | -8 |
| **** | 24 | 13 | 18 | 35 | 94 | 53 | 194 | 78 | 333 |
| **** | 25 | 12.4 | 63 | 33 | -47 | 50 | -20 | 73 | 15 |
| **** | 26 | 13.6 | 45 | 37 | -17 | 55 | 22 | 82 | 82 |
| **** | 27 | 14 | 13 | 38 | 192 | 57 | 338 | 85 | 553 |
| **** | 28 | 12.7 | 16 | 34 | 112 | 51 | 218 | 76 | 375 |
| **** | 29 | 12.8 | 16 | 34 | 112 | 52 | 225 | 76 | 375 |
| **** | 30 | 12.2 | 34 | 33 | -2 | 49 | 44 | 72 | 111 |

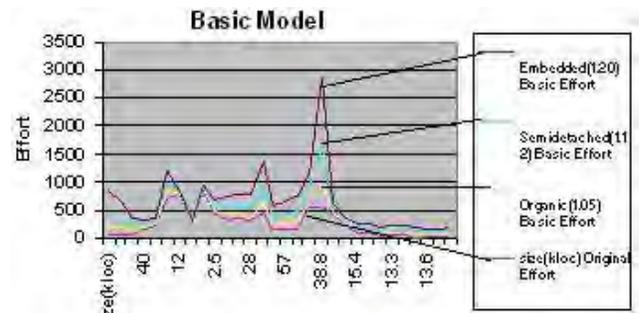

Fig. 1 shows COCOMO81 Basic model graph. Original effort is below for all the possibilities of calculated effort.

TABLE II
COCOMO81 Intermediate Model with nominal values

| References | project no. | Size(kloc) | Original Effort | Organic(1.05) Interm Effort | Error % | Semidetached(1.12) Interm Effort | Error % | Embedded(1.20) Interm Effort | Error % |
|---|---|---|---|---|---|---|---|---|---|
| * | 1 | 50 | 47 | 194 | 312 | 239 | 408 | 306 | 551 |
| * | 2 | 40 | 66 | 153 | 131 | 186 | 181 | 234 | 254 |
| * | 3 | 22 | 66 | 82 | 24 | 95 | 43 | 114 | 72 |
| * | 4 | 13 | 159 | 47 | -70 | 53 | -66 | 60 | -62 |
| * | 5 | 12 | 218 | 43 | -80 | 48 | -77 | 55 | -74 |
| * | 6 | 34 | 723 | 129 | -82 | 155 | -78 | 192 | -73 |
| * | 7 | 6.2 | 775 | 21 | -97 | 23 | -97 | 25 | -96 |
| * | 8 | 2.5 | 312 | 8 | -97 | 8 | -97 | 8 | -97 |
| * | 9 | 5.3 | 883 | 18 | -97 | 19 | -97 | 20 | -97 |
| * | 10 | 19.5 | 433 | 72 | -83 | 83 | -80 | 98 | -77 |
| * | 11 | 28 | 337 | 105 | -68 | 125 | -62 | 152 | -54 |
| * | 12 | 30 | 345 | 113 | -67 | 135 | -60 | 165 | -52 |
| * | 13 | 32 | 302 | 121 | -59 | 145 | -51 | 179 | -40 |
| * | 14 | 57 | 452 | 223 | -50 | 277 | -38 | 358 | -20 |
| ** | 15 | 30.8 | 143.7 | 116 | -19 | 139 | -3 | 171 | 18 |
| ** | 16 | 34.8 | 161.3 | 132 | -18 | 159 | -1 | 198 | 22 |
| ** | 17 | 38.8 | 178.6 | 149 | -16 | 180 | 0 | 225 | 25 |
| *** | 18 | 39 | 542 | 149 | -72 | 181 | -66 | 227 | -58 |
| **** | 19 | 128.6 | 557 | 524 | -5 | 690 | 23 | 951 | 70 |
| **** | 20 | 15.4 | 400 | 56 | -86 | 64 | -84 | 74 | -81 |
| **** | 21 | 11.3 | 240 | 40 | -83 | 45 | -81 | 51 | -78 |
| **** | 22 | 12.3 | 95 | 44 | -53 | 49 | -48 | 56 | -41 |
| **** | 23 | 13.3 | 87 | 48 | -44 | 54 | -37 | 62 | -28 |
| **** | 24 | 13 | 18 | 47 | 161 | 53 | 194 | 60 | 233 |
| **** | 25 | 12.4 | 63 | 45 | -28 | 50 | -20 | 57 | -9 |
| **** | 26 | 13.6 | 45 | 49 | 8 | 55 | 22 | 64 | 42 |
| **** | 27 | 14 | 13 | 51 | 292 | 57 | 338 | 66 | 407 |
| **** | 28 | 12.7 | 16 | 46 | 187 | 51 | 218 | 59 | 268 |
| **** | 29 | 12.8 | 16 | 46 | 187 | 52 | 225 | 59 | 268 |
| **** | 30 | 12.2 | 34 | 44 | 29 | 49 | 44 | 56 | 64 |

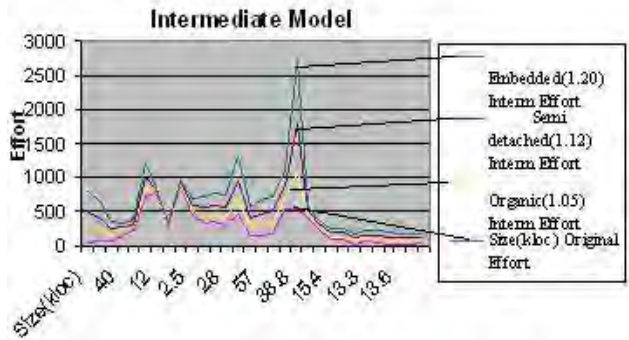

Fig. 2 shows COCOMO81 Intermediate model graph with nominal values.

TABLE III
COCOMO81 Intermediate Model with High values

| References | Project no. | Size(kloc) | Original Effort | Organic(1.05) | | Semi-detached(1.12) | | Embedded(1.20) | |
|---|---|---|---|---|---|---|---|---|---|
| | | | | Interm Effort | Error % | Interm Effort | Error % | Interm Effort | Error % |
| * | 1 | 50 | 47 | 199 | 323 | 245 | 421 | 313 | 565 |
| * | 2 | 40 | 66 | 157 | 137 | 191 | 189 | 240 | 263 |
| * | 3 | 22 | 66 | 84 | 27 | 98 | 48 | 117 | 77 |
| * | 4 | 13 | 159 | 48 | -69 | 54 | -66 | 62 | -61 |
| * | 5 | 12 | 218 | 44 | -79 | 49 | -77 | 56 | -74 |
| * | 6 | 34 | 723 | 133 | -81 | 159 | -78 | 197 | -72 |
| * | 7 | 6.2 | 775 | 22 | -97 | 23 | -97 | 25 | -96 |
| * | 8 | 2.5 | 312 | 8 | -97 | 8 | -97 | 8 | -97 |
| * | 9 | 5.3 | 883 | 18 | -97 | 19 | -97 | 21 | -97 |
| * | 10 | 19.5 | 433 | 74 | -82 | 85 | -80 | 101 | -76 |
| * | 11 | 28 | 337 | 108 | -67 | 128 | -62 | 156 | -53 |
| * | 12 | 30 | 345 | 116 | -66 | 138 | -60 | 169 | -51 |
| * | 13 | 32 | 302 | 124 | -58 | 149 | -50 | 183 | -39 |
| * | 14 | 57 | 452 | 228 | -49 | 284 | -37 | 367 | -18 |
| ** | 15 | 30.8 | 143.7 | 119 | -17 | 142 | -1 | 175 | 21 |
| ** | 16 | 34.8 | 161.3 | 136 | -15 | 163 | 1 | 203 | 25 |
| ** | 17 | 38.8 | 178.6 | 152 | -14 | 185 | 3 | 231 | 29 |
| *** | 18 | 39 | 542 | 153 | -71 | 186 | -65 | 232 | -57 |
| **** | 19 | 128.6 | 557 | 537 | -3 | 708 | 27 | 974 | 74 |
| **** | 20 | 15.4 | 400 | 57 | -85 | 65 | -83 | 76 | -81 |
| **** | 21 | 11.3 | 240 | 41 | -82 | 46 | -80 | 52 | -78 |
| **** | 22 | 12.3 | 95 | 45 | -52 | 51 | -46 | 58 | -38 |
| **** | 23 | 13.3 | 87 | 49 | -43 | 55 | -36 | 64 | -26 |
| **** | 24 | 13 | 18 | 48 | 166 | 54 | 200 | 62 | 244 |
| **** | 25 | 12.4 | 63 | 46 | -26 | 51 | -19 | 58 | -7 |
| **** | 26 | 13.6 | 45 | 50 | 11 | 57 | 26 | 65 | 44 |
| **** | 27 | 14 | 13 | 52 | 300 | 59 | 353 | 68 | 423 |
| **** | 28 | 12.7 | 16 | 47 | 193 | 52 | 225 | 60 | 275 |
| **** | 29 | 12.8 | 16 | 47 | 193 | 53 | 231 | 61 | 281 |
| **** | 30 | 12.2 | 34 | 45 | 32 | 50 | 47 | 57 | 67 |

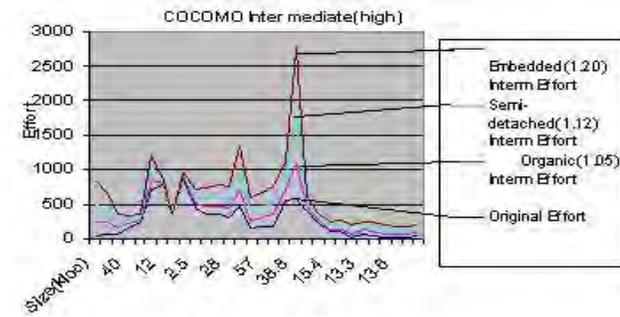

Fig. 3 shows COCOMO81 Intermediate model graph with High values.

TABLE IV
COCOMO81 Detailed Model with nominal values

| References | Project no. | Size(kloc) | Original Effort | Organic(1.05)(nominal) | | Semi-detached(1.12) | | Embedded(1.20) | |
|---|---|---|---|---|---|---|---|---|---|
| | | | | Effort | Error % | Effort | Error % | Effort | Error % |
| * | 1 | 50 | 47 | 206 | 338 | 256 | 444 | 330 | 602 |
| * | 2 | 40 | 66 | 163 | 146 | 199 | 201 | 252 | 281 |
| * | 3 | 22 | 66 | 87 | 31 | 102 | 54 | 123 | 86 |
| * | 4 | 13 | 159 | 50 | -68 | 56 | -64 | 65 | -59 |
| * | 5 | 12 | 218 | 46 | -78 | 51 | -76 | 59 | -72 |
| * | 6 | 34 | 723 | 137 | -81 | 166 | -77 | 208 | -71 |
| * | 7 | 6.2 | 775 | 23 | -97 | 24 | -96 | 27 | -96 |
| * | 8 | 2.5 | 312 | 8 | -97 | 8 | -97 | 9 | -97 |
| * | 9 | 5.3 | 883 | 19 | -97 | 20 | -97 | 22 | -97 |
| * | 10 | 19.5 | 433 | 76 | -82 | 89 | -79 | 106 | -75 |
| * | 11 | 28 | 337 | 112 | -66 | 134 | -60 | 164 | -51 |
| * | 12 | 30 | 345 | 120 | -65 | 144 | -58 | 179 | -48 |
| * | 13 | 32 | 302 | 129 | -57 | 155 | -48 | 193 | -36 |
| * | 14 | 57 | 452 | 236 | -47 | 297 | -34 | 386 | -14 |
| ** | 15 | 30.8 | 143.7 | 124 | -13 | 149 | 3 | 184 | 28 |
| ** | 16 | 34.8 | 161.3 | 140 | -13 | 171 | 6 | 214 | 32 |
| ** | 17 | 38.8 | 178.6 | 158 | -11 | 193 | 8 | 243 | 36 |
| *** | 18 | 39 | 542 | 158 | -70 | 194 | -64 | 245 | -54 |
| **** | 19 | 128.6 | 557 | 556 | 0 | 739 | 32 | 1027 | 84 |
| **** | 20 | 15.4 | 400 | 59 | -85 | 68 | -83 | 80 | -80 |
| **** | 21 | 11.3 | 240 | 43 | -82 | 48 | -80 | 55 | -77 |
| **** | 22 | 12.3 | 95 | 47 | -50 | 53 | -44 | 61 | -35 |
| **** | 23 | 13.3 | 87 | 51 | -41 | 58 | -33 | 67 | -22 |
| **** | 24 | 13 | 18 | 50 | 177 | 56 | 211 | 65 | 261 |
| **** | 25 | 12.4 | 63 | 47 | -25 | 53 | -15 | 62 | -1 |
| **** | 26 | 13.6 | 45 | 52 | 15 | 59 | 31 | 69 | 53 |
| **** | 27 | 14 | 13 | 54 | 315 | 61 | 369 | 71 | 446 |
| **** | 28 | 12.7 | 16 | 48 | 200 | 55 | 243 | 63 | 293 |
| **** | 29 | 12.8 | 16 | 49 | 206 | 55 | 243 | 64 | 300 |
| **** | 30 | 12.2 | 34 | 46 | 35 | 52 | 52 | 60 | 76 |

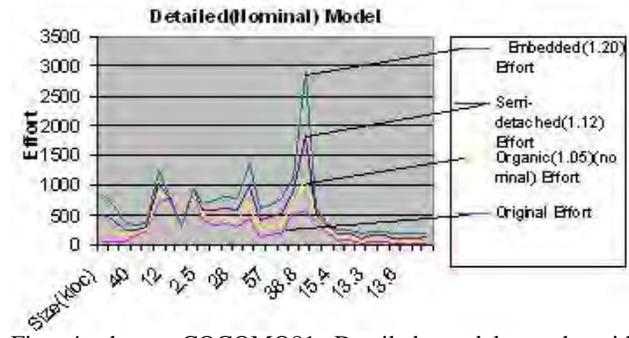

Fig. 4 shows COCOMO81 Detailed model graph with nominal values.

TABLE V
COCOMO81 Detailed Model with High values

| References | Project no. | Size(kloc) | Original Effort | Organic(1.05)(high) | | Semi-detached(1.12) | | Embedded(1.20) | |
|---|---|---|---|---|---|---|---|---|---|
| | | | | Effort | Error % | Effort | Error % | Effort | Error % |
| * | 1 | 50 | 47 | 211 | 348 | 262 | 457 | 338 | 619 |
| * | 2 | 40 | 66 | 167 | 153 | 204 | 209 | 259 | 292 |
| * | 3 | 22 | 66 | 89 | 34 | 104 | 57 | 126 | 90 |
| * | 4 | 13 | 159 | 51 | -67 | 58 | -63 | 67 | -57 |
| * | 5 | 12 | 218 | 47 | -78 | 53 | -75 | 61 | -72 |
| * | 6 | 34 | 723 | 140 | -80 | 170 | -76 | 213 | -70 |
| * | 7 | 6.2 | 775 | 23 | -97 | 25 | -96 | 27 | -96 |
| * | 8 | 2.5 | 312 | 9 | -97 | 9 | -97 | 9 | -97 |
| * | 9 | 5.3 | 883 | 20 | -97 | 21 | -97 | 22 | -97 |
| * | 10 | 19.5 | 433 | 78 | -81 | 91 | -78 | 109 | -74 |
| * | 11 | 28 | 337 | 114 | -66 | 137 | -59 | 168 | -50 |
| * | 12 | 30 | 345 | 123 | -64 | 148 | -57 | 183 | -46 |
| * | 13 | 32 | 302 | 132 | -56 | 159 | -47 | 198 | -34 |
| * | 14 | 57 | 452 | 242 | -46 | 304 | -32 | 396 | -12 |
| ** | 15 | 30.8 | 143.7 | 126 | -12 | 152 | 5 | 189 | 31 |
| ** | 16 | 34.8 | 161.3 | 144 | -10 | 175 | 8 | 219 | 35 |
| ** | 17 | 38.8 | 178.6 | 161 | -9 | 197 | 10 | 249 | 39 |
| *** | 18 | 39 | 542 | 162 | -70 | 198 | -63 | 251 | -53 |
| **** | 19 | 128.6 | 557 | 569 | 2 | 757 | 35 | 1051 | 88 |
| **** | 20 | 15.4 | 400 | 61 | -84 | 70 | -82 | 82 | -79 |
| **** | 21 | 11.3 | 240 | 44 | -81 | 49 | -79 | 56 | -76 |
| **** | 22 | 12.3 | 95 | 48 | -49 | 54 | -43 | 62 | -34 |
| **** | 23 | 13.3 | 87 | 52 | -40 | 59 | -32 | 69 | -20 |
| **** | 24 | 13 | 18 | 51 | 183 | 58 | 222 | 67 | 272 |
| **** | 25 | 12.4 | 63 | 48 | -23 | 55 | -12 | 63 | 0 |
| **** | 26 | 13.6 | 45 | 53 | 17 | 61 | 35 | 70 | 55 |
| **** | 27 | 14 | 13 | 55 | 323 | 63 | 384 | 73 | 461 |
| **** | 28 | 12.7 | 16 | 50 | 212 | 56 | 250 | 65 | 306 |
| **** | 29 | 12.8 | 16 | 50 | 212 | 57 | 256 | 65 | 306 |
| **** | 30 | 12.2 | 34 | 48 | 41 | 54 | 58 | 62 | 82 |

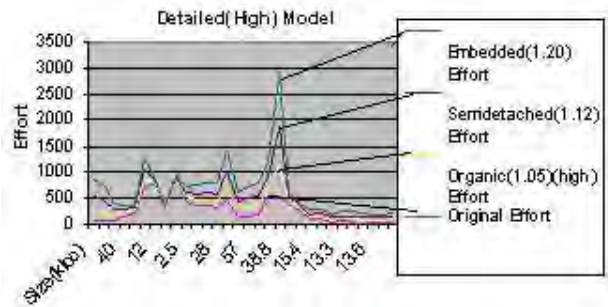

Fig. 5 shows COCOMO81 Detailed model graph with High values.

TABLE VI
COCOMOII Early Design Model

| References | Project no. | Size(kloc) | Original Effort | Earlydesign(Nominal) Effort | Error % | Earlydesign(high) Effort | Error % |
|---|---|---|---|---|---|---|---|
| * | 1 | 50 | 47 | 122 | 159 | 138 | 193 |
| * | 2 | 40 | 66 | 98 | 48 | 110 | 66 |
| * | 3 | 22 | 66 | 53 | -19 | 60 | -9 |
| * | 4 | 13 | 159 | 31 | -80 | 35 | -77 |
| * | 5 | 12 | 218 | 29 | -86 | 33 | -84 |
| * | 6 | 34 | 723 | 83 | -88 | 93 | -87 |
| * | 7 | 6.2 | 775 | 15 | -98 | 17 | -97 |
| * | 8 | 2.5 | 312 | 6 | -98 | 6 | -98 |
| * | 9 | 5.3 | 883 | 12 | -98 | 14 | -98 |
| * | 10 | 19.5 | 433 | 47 | -89 | 53 | -87 |
| * | 11 | 28 | 337 | 68 | -79 | 77 | -77 |
| * | 12 | 30 | 345 | 73 | -78 | 82 | -76 |
| * | 13 | 32 | 302 | 78 | -74 | 88 | -70 |
| * | 14 | 57 | 452 | 139 | -69 | 157 | -65 |
| ** | 15 | 30.8 | 143.7 | 75 | -47 | 85 | -40 |
| ** | 16 | 34.8 | 161.3 | 85 | -47 | 96 | -40 |
| ** | 17 | 38.8 | 178.6 | 95 | -46 | 107 | -40 |
| *** | 18 | 39 | 542 | 95 | -82 | 107 | -80 |
| **** | 19 | 128.6 | 557 | 315 | -43 | 355 | -36 |
| **** | 20 | 15.4 | 400 | 37 | -90 | 42 | -89 |
| **** | 21 | 11.3 | 240 | 27 | -88 | 31 | -87 |
| **** | 22 | 12.3 | 95 | 30 | -68 | 33 | -65 |
| **** | 23 | 13.3 | 87 | 32 | -63 | 36 | -58 |
| **** | 24 | 13 | 18 | 31 | 72 | 35 | 94 |
| **** | 25 | 12.4 | 63 | 30 | -52 | 34 | -46 |
| **** | 26 | 13.6 | 45 | 33 | -26 | 37 | -17 |
| **** | 27 | 14 | 13 | 34 | 161 | 38 | 192 |
| **** | 28 | 12.7 | 16 | 31 | 93 | 35 | 118 |
| **** | 29 | 12.8 | 16 | 31 | 93 | 35 | 118 |
| **** | 30 | 12.2 | 34 | 29 | -14 | 33 | -2 |

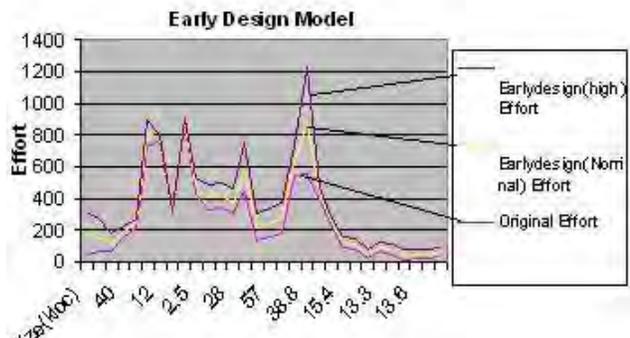

Fig. 6 shows graph for COCOMOII Early Design model.

TABLE VII
COCOMOII Post Architecture Model

| References | project no. | size(kloc) | Original Effort | postarch(n-n) Effort | Error% | postarch(n-high) Effort | Error% | postarch(high-n) Effort | Error% | postarch(h-h) Effort | Error % |
|---|---|---|---|---|---|---|---|---|---|---|---|
| * | 1 | 50 | 47 | 229 | 387 | 248 | 427 | 188 | 300 | 282 | 500 |
| * | 2 | 40 | 66 | 177 | 168 | 192 | 190 | 147 | 122 | 218 | 230 |
| * | 3 | 22 | 66 | 89 | 34 | 96 | 45 | 76 | 15 | 109 | 65 |
| * | 4 | 13 | 159 | 48 | -69 | 52 | -67 | 42 | -73 | 59 | -62 |
| * | 5 | 12 | 218 | 44 | -79 | 48 | -77 | 39 | -82 | 54 | -75 |
| * | 6 | 34 | 723 | 147 | -79 | 159 | -78 | 123 | -82 | 180 | -75 |
| * | 7 | 6.2 | 775 | 20 | -97 | 22 | -97 | 18 | -97 | 25 | -96 |
| * | 8 | 2.5 | 312 | 7 | -97 | 7 | -97 | 6 | -98 | 8 | -97 |
| * | 9 | 5.3 | 883 | 17 | -98 | 18 | -97 | 15 | -98 | 21 | -97 |
| * | 10 | 19.5 | 433 | 77 | -82 | 84 | -80 | 66 | -84 | 95 | -78 |
| * | 11 | 28 | 337 | 117 | -65 | 127 | -62 | 99 | -70 | 144 | -57 |
| * | 12 | 30 | 345 | 127 | -63 | 138 | -60 | 107 | -68 | 156 | -54 |
| * | 13 | 32 | 302 | 137 | -54 | 148 | -50 | 115 | -61 | 168 | -44 |
| * | 14 | 57 | 452 | 266 | -41 | 289 | -36 | 217 | -51 | 327 | -27 |
| ** | 15 | 30.8 | 143.7 | 131 | -8 | 142 | -1 | 110 | -23 | 161 | 12 |
| ** | 16 | 34.8 | 161.3 | 151 | -6 | 164 | 1 | 126 | -21 | 185 | 14 |
| ** | 17 | 38.8 | 178.6 | 171 | -4 | 185 | 3 | 142 | -20 | 210 | 17 |
| *** | 18 | 39 | 542 | 172 | -68 | 186 | -65 | 143 | -73 | 211 | -61 |
| **** | 19 | 128.6 | 557 | 679 | 21 | 737 | 32 | 532 | -4 | 835 | 49 |
| **** | 20 | 15.4 | 400 | 59 | -85 | 64 | -84 | 51 | -87 | 72 | -82 |
| **** | 21 | 11.3 | 240 | 41 | -82 | 44 | -81 | 36 | -85 | 50 | -79 |
| **** | 22 | 12.3 | 95 | 45 | -52 | 49 | -48 | 40 | -57 | 56 | -41 |
| **** | 23 | 13.3 | 87 | 49 | -43 | 54 | -37 | 43 | -50 | 61 | -29 |
| **** | 24 | 13 | 18 | 48 | 166 | 52 | 188 | 42 | 133 | 59 | 227 |
| **** | 25 | 12.4 | 63 | 46 | -26 | 50 | -20 | 40 | -36 | 56 | -11 |
| **** | 26 | 13.6 | 45 | 51 | 13 | 55 | 22 | 45 | 0 | 63 | 40 |
| **** | 27 | 14 | 13 | 53 | 307 | 57 | 338 | 46 | 253 | 65 | 400 |
| **** | 28 | 12.7 | 16 | 47 | 193 | 51 | 218 | 41 | 156 | 58 | 262 |
| **** | 29 | 12.8 | 16 | 47 | 193 | 51 | 218 | 42 | 162 | 58 | 262 |
| **** | 30 | 12.2 | 34 | 45 | 32 | 49 | 44 | 39 | 14 | 55 | 61 |

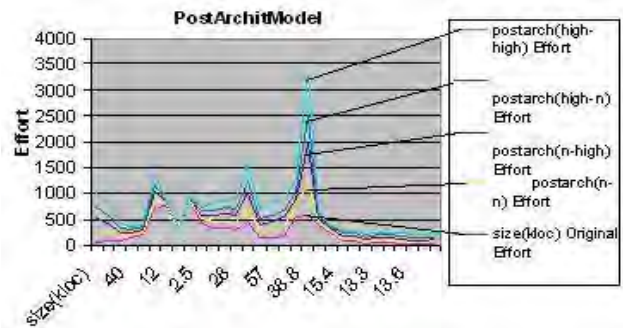

Fig. 7

Fig. 7 shows COCOMOII Post Architecture model graph.

Similarly the variations are found in SLIM, FP model. [11]

VI. DISCUSSION AND FUTURE WORK

This research would emphasize on the salience of software reuse principles in cognition with software cost estimation. Also we try to articulate the multifarious ways in which software reuse aids the cause of cost estimation. We are speculating on devising a cost estimation model which highlights the preponderance of reuse and else reduce the MRE. We are also trying to peg the different forms of reuse to reduce cost and MRE.

VII. CONCLUSION

This work on Software Cost Estimation explores contemporary cost estimation models and different ways of guesstimating the cost. It compares COCOMOII which has been widely used over the past years. Also it ponders over the other models SLIM, Function Point Model and Delphi which have had profound influence especially in practice. Furthermore it attempts to put forth the gist of software reuse for future use.

c